\documentclass[aps,pra,superscriptaddress,twocolumn,a4paper,reprint,eprint,longbibliography,footinbib,floatfix]{revtex4-2}
\pdfoutput=1
\usepackage[percent]{overpic}
\usepackage{physics}
\usepackage[utf8]{inputenc}
\usepackage{graphicx}
\usepackage{amsmath}
\usepackage{amsfonts}
\usepackage{amssymb}
\usepackage{bbm}
\usepackage{bm}
\usepackage{bbold}
\usepackage[usenames,dvipsnames,svgnames,table]{xcolor}
\usepackage[USenglish]{babel}
\usepackage{hyperref}
\usepackage{grffile}
\usepackage{multirow}
\usepackage{hhline}

\newcommand{\ignore}[1]{}

\allowdisplaybreaks

\pdfpageattr{/Group <</S /Transparency /I true /CS /DeviceRGB>>}

\begin{document}

\title{Pair-breaking in superconductors with strong
  spin-orbit coupling} 

\author{D. C. Cavanagh}
\affiliation{Department of Physics, University of Otago, P.O. Box
56, Dunedin 9054, New Zealand}
\author{Daniel F. Agterberg}
\email{agterber@uwm.edu}
\affiliation{Department of Physics, University of Wisconsin, Milwaukee, Wisconsin 53201, USA}
\author{P. M. R. Brydon}
\email{philip.brydon@otago.ac.nz}
\affiliation{Department of Physics and MacDiarmid Institute for
Advanced Materials and Nanotechnology, University of Otago, P.O. Box
56, Dunedin 9054, New Zealand}

\date{July 4, 2022}

\begin{abstract}
We study the influence of symmetry-breaking perturbations on
superconductivity in multiorbital materials, with a particular focus
on an external magnetic field. We introduce the field-fitness function
which characterizes the pair-breaking effects of the perturbation on a
given superconducting state. For even parity superconductors we find
that this field-fitness function for an external magnetic field is
one, implying that the paramagnetic response is controlled only by a
generalized effective $g$-factor. For odd parity superconductors, the
interplay of the effective $g$-factor and the field-fitness function
can lead to counter-intuitive results. We demonstrate this for
$p$-wave pairing in the effective $j=3/2$ electronic
states of the Luttinger-Kohn model. 
\end{abstract}
\maketitle

{\it Introduction.} A diverse variety of 
  superconductors have recently been found to exhibit critical fields 
  far  exceeding the Pauli limiting field, e.g. UTe$_2$ \cite{Ran2019},
  CeRh$_2$As$_2$ \cite{Khim2021}, UCoGe\cite{Aoki2019},
  URhGe\cite{Aoki2019}, and YbRh$_2$Si$_2$ \cite{Nguyen2021}. In some
  of these materials superconductivity even appears as a re-entrant
  phase \cite{Aoki2019,Ran2019}, again far above the Pauli field. The
  origin of this anomalous high-field behaviour has been
  attributed to spin-triplet superconductivity. However, the
  fermiology of these materials is complicated, with multiple bands
  crossing the Fermi surface.  Moreover, spin-orbit coupling is
  expected to be large due to the presence of heavy elements. Due to
  the interplay of the normal-state band structure, the structure of
  the odd-parity pairing potential, and the applied magnetic field, it
  is not clear that the established theory for the magnetic
  response of a triplet superconductor is applicable~\cite{Mineev1999,Sigrist2005Intro}. 

Concurrent to these experimental developments, it has been realized
that internal degrees of freedom of the band electrons,
e.g. sublattice or orbital, can profoundly impact the magnetic
response of even-parity superconductors. In particular, the large critical
fields observed in artificial Rashba 
heterostructures~\cite{Shimozawa2014, Watanabe2015},
CeRh$_2$As$_2$~\cite{Khim2021,ce122_theory_nonsym}, and
WTe$_2$~\cite{Xie_2020} are believed to arise from a 
  ``hidden'' antisymmetric spin-orbit coupling (ASOC)~\cite{Zhang2014}.
  This ASOC is odd in momentum and has opposite signs for
  internal degrees of freedom that are related by inversion symmetry
  (IS). This preserves IS and the two-fold degeneracy
  of the band electron states. Similar to
  noncentrosymmetric materials,~\cite{smidman2017} however, the ASOC
  reduces the Zeeman splitting of the band states and so enhances the
  Pauli limit~\cite{youn:2012}. For 
  odd-parity superconductivity, however, the effect of the ASOC on the
  critical fields is not as well
  explored. However, in one remarkable example, CeRh$_2$As$_2$,
  spin-singlet pairing interactions give way to extremely high
  critical fields due to the formation of an odd-parity
  superconducting state that is stabilized by the
  ASOC~\cite{Yoshida2012,Khim2021}.   

There also exist materials where a symmetric spin-orbit coupling
(SSOC), i.e. the spin-orbit coupling is even in momentum, is
important. This influence of this SSOC on superconductivity has been
studied in the context of the iron pnictides~\cite{Vafek2017}, Sr$_2$RuO$_4$~\cite{Ramires2016,SRO},
half-Heusler materials,~\cite{Brydon2016,Savary2017} and the
pyrochlore lattice~\cite{pyrochlore}. The latter two cases support 
effective $j=3/2$ electronic states which exhibit properties that are
quite different from the more usual $j=1/2$
states~\cite{Venderbos2018}. The magnetic response of
even-parity superconducting states in such materials shows similar
features to the case of the ASOC~\cite{Kim2022}; the response of
odd-parity 
superconductivity also remains poorly understood. Indeed, one of the main
results of this work is to reveal the counter-intuitive response of
odd-parity states in $j=3/2$ materials to applied fields.  

In this article we examine the influence of the spin-orbit coupling on
the response of a superconducting state to a perturbation which breaks
either IS or time-reversal symmetry (TRS), with a focus on the
familiar example of an applied magnetic field. 
Within a general minimal model for systems with both antisymmetric
and symmetric spin-orbit coupling, we 
show that the response of the superconducting state to the perturbation is
fully determined by two basis-independent quantities: a generalized
effective $g$-factor, and a parameter that quantifies the
pair-breaking due to the field, which we term the field-fitness
function $F_h$ by analogy with the superconducting fitness
\cite{Ramires2016,Ramires2018}. In the case of an external magnetic
field, these quantities also control the spin susceptibility in the
superconducting state. For even-parity superconductors,
$F_h=1$, so that the response is given solely by the effective
$g$-factor; in contrast, odd-parity superconducting states display a
complicated interplay of the field-fitness and the effective
$g$-factor. We apply our general theory to odd-parity
superconductivity in $j=3/2$ materials, where we find that the SSOC
leads to a magnetic response very unlike that of $j=1/2$
superconductors.

{\it General theory}. We consider a system described by the Bogoliubov-de Gennes Hamiltonian
\begin{equation}
H=\sum_{\bm{k}}\vec{\Psi}_{\bm{k}}^{\dagger}\begin{pmatrix}
\mathcal{H}_{0,\bm{k}}  & \Delta_{\bm{k}}\\
\Delta_{\bm{k}}^\dagger & -\mathcal{H}_{0,-\bm{k}}^{T} 
\end{pmatrix}\vec{\Psi}_{\bm{k}},\label{eq:BdG_Ham}
\end{equation}
where $\vec{\Psi}_{\bm{k}} = (\vec{c}_{\bm k},\vec{c}^\dagger_{-\bm
  k})$, with $\vec{c}_{\bm k}$ representing a spinor of
annihilation operators for fermions with four internal
degrees of freedom.
In the presence of both TRS and IS, the most general form of the
Hamiltonian matrix ${\cal H}_{0,\bm{k}}$
is~\cite{Brydon2018b,Abrikosov1974}
\begin{equation}
\mathcal{H}_{0,\bm{k}}=\varepsilon_{0,\bm{k}}\mathbb{1}+\vec{\varepsilon}_{\bm{k}}\cdot\vec{\gamma}=\varepsilon_{0,\bm{k}}\mathbb{1}+\bar{\mathcal{H}}_{0,\bm{k}},\label{eq:H0}
\end{equation}
where $\vec{\gamma}=(\gamma^1, \gamma^2, \gamma^3, \gamma^4, \gamma^5
)$ are the mutually anticommuting Euclidean Dirac matrices 
with coefficients $\vec{\varepsilon}_{\bm k}=(\varepsilon_{1,\bm{k}},
\varepsilon_{2,\bm{k}}, \varepsilon_{3,\bm{k}},
\varepsilon_{4,\bm{k}}, \varepsilon_{5,\bm{k}} )$. The Hamiltonian has
doubly-degenerate eigenenergies $\xi_{\pm,{\bm k}}=\varepsilon_{0,\bm{k}}
\pm|\vec{\varepsilon}_{\bm{k}}|$, where $|\vec{\varepsilon}_{\bm{k}}| =
\sqrt{\sum_{l=1}^5\varepsilon_{l,\bm{k}}^2}$. In Eq.~\ref{eq:H0} we
have introduced $\bar{\mathcal{H}}_{0,\bm{k}} =
\vec{\epsilon}_{\bm k}\cdot\vec{\gamma}$ to denote the part of 
the Hamiltonian which depends nontrivially on the internal degrees of
freedom. 

Hamiltonians with the form of Eq.~\ref{eq:H0} 
describe a diverse range of two-band
systems~\cite{FuBerg2010,Yanase2016,Vafek2017,Agterberg2017FeSe,Yoshida2012,Ilic2017,Brydon2016,Brydon2018b,Xie_2020},
and so
the exact form of the $\gamma$ matrices depends on the system under
consideration~\cite{Denys2021}. For
spin-$\frac{1}{2}$ systems, we generally construct the $\gamma$
matrices as Kronecker products of Pauli matrices acting in the orbital
and spin spaces. In this case we choose $\gamma^1$ and  $\gamma^2$ to
be trivial in the spin space, with these contributions to the
Hamiltonian describing purely orbital effects, while the remaining
matrices couple the spin and orbital degrees of freedom,
  accounting for SOC in the system.  As we shall see, 
this parameterization is also valid for spin-$\frac{3}{2}$ systems,
even though the underlying orbital
  and spin degrees of freedom cannot be factorized.

The pairing potential appearing in
Eq. \ref{eq:BdG_Ham} is written  $\Delta_{\bm k} =
\Delta_0 \tilde{\Delta}_{{\bm k}}U_T$, where $\Delta_0$ is the magnitude,
$\tilde{\Delta}_{\bm k}$ encodes the dependence on the momentum and
the internal degrees of freedom, and $U_T$ is the unitary part of the
time-reversal operator. The general form for even- ($e$) and odd-parity
($o$) states is 
\begin{align}
\tilde{\Delta}^{(e)}_{\bm k} &= \sum_{a=0}^5e^a_{\bm{k}}\gamma^{a} \label{eq:genDelta_e}
                               \\
\tilde{\Delta}^{(o)}_{\bm k} &= \sum_{a=1}^4\sum_{b>a}o^{ab}_{\bm{k}}i\gamma^{a}\gamma^{b}\label{eq:genDelta_o}
\end{align}
where $e^a_{\bm{k}}$ and $o^{ab}_{\bm k}$ are normalized form factors.
Note that
only when the internal degrees of freedom transform trivially under
inversion are 
the functions $e^a_{\bm{k}}$ and $o^{ab}_{\bm k}$ necessarily
even and odd in momentum, respectively.

Due to the mixing of orbital and spin, 
the pairing potential in the band basis typically has 
both intraband and interband matrix elements. The intraband gap is
particularly important as it is responsible for the Cooper
instability. Assuming that $\Delta_0$ is small compared to the band
separation, the gap in band $a$ is given by 
\begin{equation}
\left|\Delta_{a,\bm{k}}\right|^2=\Delta_0^2\frac{\text{Tr}
 \lbrace |\{
      \bar{\mathcal{H}}_{0,\bm{k}},\tilde{\Delta}_{\bm{k}}\}
|^2\mathcal{P}_{a,{\bm k}}\rbrace}{8\left|\vec{\varepsilon}_{\bm{k}}\right|^2} ,\label{eq:FitA_1}
\end{equation}
 where $\{\bar{\mathcal{H}}_{0,\bm{k}},\tilde{\Delta}_{\bm{k}}\}U_T = F_A$
is the superconducting fitness as defined in
Refs. \cite{Ramires2016,Ramires2018}, and $\mathcal{P}_{a,{\bm k}}=
\frac{1}{2}(\mathbb{1} +
a\bar{\mathcal{H}}_{0,\bm{k}}/|\vec{\epsilon}|)$ projects into the
$a=\pm$ band. The  projection operator is  necessary to account for band-dependence of
the intraband pairing, which can arise when $\tilde{\Delta}_{\bm
  k}\tilde{\Delta}_{\bm k}^\dagger \not\propto \mathbb{1}$. 

To investigate the effect of  symmetry-breaking, we introduce the
perturbation Hamiltonian 
\begin{equation}
\delta H=\sum_{\bm{k}}\vec{\Psi}_{\bm{k}}^{T}\begin{pmatrix}
 \mathcal{H}_{h,\bm{k}} & 0\\
0 & - \mathcal{H}_{h,-\bm{k}}^{T}
\end{pmatrix}\vec{\Psi}_{\bm{k}}.\label{eq:BdG_pertHam}
\end{equation}
We adopt a general form of $\mathcal{H}_{h,\bm{k}}$ 
\begin{equation}
\mathcal{H}_{h,\bm{k}} = \sum_{\alpha=0}^5
h_{0\alpha,\bm{k}}\gamma^{\alpha} + \sum_{\alpha=1}^4\sum_{\beta>\alpha} h_{\alpha\beta,\bm{k}}i\gamma^{\alpha} \gamma^{\beta},\label{eq:H_h}
\end{equation}
 The perturbation lifts the twofold
degeneracy of the normal state spectrum:
For sufficiently well-separated bands, the perturbed energies of band
$a$ are $\xi_{a,\pm,{\bm k}}\approx \xi_{a,{\bm k}} \pm \tilde{g}_{a,{\bm k}} h_{\bm k}$, where $h^2_{\bm k}=\sum_{\alpha,\beta} h_{\alpha\beta,\bm{k}}^2
=\text{Tr}\{|\mathcal{H}_{h,\bm{k}}|^2\}/4$, and the effective
$g$-factor in band $a$ is
\begin{equation}
\tilde{g}_{a,{\bm k}}^2 = \frac{\text{Tr}\lbrace
    |\lbrace
          \bar{\mathcal{H}}_{0,\bm{k}},\mathcal{H}_{h,\bm{k}}\rbrace|^2\mathcal{P}_{a,{\bm k}}
  \rbrace}{8\left|\vec{\varepsilon}_{\bm{k}}\right|^2h^2_{\bm k}}.\label{eq:h_Fit}
\end{equation}
Equation~\ref{eq:h_Fit} resembles the
expression for the intraband superconducting gap Eq.~\ref{eq:FitA_1},
and can be similarly interpreted as giving the splitting of the bands due
to the projection of the perturbation onto band $a$. When $|\mathcal{H}_{h,\bm{k}}|^2\propto
\mathbb{1}$ the splitting of 
the energy spectrum is independent of the band index,
i.e. $g_{a,\bm{k}}=g_{\bm{k}}$; more generally, when  $|\mathcal{H}_{h,\bm{k}}|^2\not\propto
\mathbb{1}$ the effective $g$-factors are different in each band,
which is accounted for by the projection
operator in Eq.~\ref{eq:h_Fit}.
For definiteness, in the following we
consider only perturbations which break TRS but preserve IS; the
results for perturbations which break IS but preserve TRS are similar
and provided in appendix~\ref{App:ISB}.

{\it Pair-breaking.} The lifting of the two-fold degeneracy of the band states by the
perturbation  generally suppresses the superconductivity.
The central result of our work is that the pair-breaking
effects of the perturbation in band 
$a$ can be quantified by the {\it field-fitness function}
\begin{equation}
	\tilde{F}_{h,{\bm k}}^{(a)}=\frac{\text{Tr}\lbrace
            |\lbrace\lbrace\bar{\mathcal{H}}_{0,\bm{k}},\tilde{\Delta}_{\bm
              k}\rbrace,\lbrace
            \bar{\mathcal{H}}_{0,\bm{k}},\mathcal{H}_{h,\bm{k}}\rbrace\rbrace|^2
            \mathcal{P}_{a,{\bm k}}
          \rbrace}{2\text{Tr}\lbrace |\lbrace
            \bar{\mathcal{H}}_{0,\bm{k}},\mathcal{H}_{h,\bm{k}}\rbrace|^2\mathcal{P}_{a,{\bm k}}
          \rbrace\text{Tr}\lbrace |\lbrace
            \bar{\mathcal{H}}_{0,\bm{k}},\tilde{\Delta}_{\bm k}\rbrace|^2\mathcal{P}_{a,{\bm
              k}}\rbrace}\,. \label{eq:FieldFitnessFunction_1}
	\end{equation}
The field-fitness function ranges in value from zero to one.
For $\tilde{F}^{(a)}_{h,{\bm k}}=0$,  the states at ${\bm k}$ and
$-{\bm k}$ involved in the intraband pairing remain degenerate, 
and so there is no pair-breaking effect. On the other
hand, the perturbation is maximally pair-breaking for
$\tilde{F}^{(a)}_{h,{\bm k}}=1$, i.e. the states paired by the
intraband pairing potential are split by the perturbation. 
An intermediate value $0<\tilde{F}^{(a)}_{h,{\bm k}}<1$ indicates that
the intraband pairing potential pairs electrons in a superposition of the
perturbed states, and there will be some pair-breaking effect.
Inserting Eq.~\ref{eq:genDelta_e} into
  Eq.~\ref{eq:FieldFitnessFunction_1}, we find that for any
  even-parity state the field fitness
  $\tilde{F}_{h,{\bm k}}^{(a)}=1$, as the numerator can be factored to
  give the denominator.  Since even-parity superconductors always pair
  time-reversed partners within the same band, any TRS-breaking
  perturbation is maximally pair-breaking.
On the other hand, odd-parity superconducting states do not
necessarily pair time-reversed states in the same band, and so they may
experience less or no
pair-breaking due to broken TRS, i.e. $0\leq \tilde{F}_{h}^{(a)} \leq 1$.

Solving the linearized gap equation in the presence of the
TRS-breaking perturbation gives the critical temperature $T_c$ in
terms of the unperturbed value $T_{c,0}$,  
\begin{widetext}
	
	\begin{equation}
	\log\left(\frac{T_c}{T_{c,0}}\right) = \sum_{a=\pm}\left\langle
          \left[\frac{\mathcal{D}_{a,\bm{k}}\left|\Delta_{a,\bm{k}}\right|^2}{\sum_{a'=\pm}\langle\mathcal{D}_{a',\bm{k}'}\left|\Delta_{a',\bm{k}'}\right|^2\rangle_{a'}}\right]
          \tilde{F}_{h,{\bm k}}^{(a)}\text{Re}\left\lbrace
            \psi\left(\frac{1}{2}\right) -
            \psi\left(\frac{1}{2}+i\frac{\tilde{g}_{a,{\bm k}}h_{\bm k}}{2\pi k_BT_c}\right)\right\rbrace\right\rangle_{a},\label{eq:logTc_h_1}
	\end{equation}
	
\end{widetext}
where $\psi(x)$ is the digamma function, $\langle\ldots\rangle_{a}$
indicates the average over the Fermi surface of band $a$,
$\mathcal{D}_{a,\bm{k}} = |\vec{\nabla}_{\bm{k}} \xi_{a,\bm{k}}|^{-1}$, 
and the
factor in the square brackets defines the fraction of the total
condensation energy due to the gap on  each band. 
The suppression of the critical temperature by a
TRS-breaking perturbation is controlled by both the field fitness
function and the effective $g$-factor, which tune the degree of
pair-breaking and the magnitude of the band splitting,
respectively. 
A brief derivation of Eq.~\ref{eq:logTc_h_1} is presented in appendix~\ref{App:general}.

{\it Magnetic susceptibility}.  We now turn to the important case where the perturbation is an
  applied magnetic field, which couples to the electron states via  the Zeeman effect. A key
experimental quantity is the magnetic susceptibility,  which in a
multiband system can be divided into
components 
due to intraband (`Pauli') and interband (`van Vleck')
transitions. The  latter is negligibly affected by superconductivity,
as the pairing potential is typically much smaller than the band
separation. On the other hand, the Pauli contribution
carries clear signatures of the pair-breaking effect.  For a field
applied along the $i$-axis, the Pauli susceptibility below the
critical temperature is given by
\begin{equation}
\chi_{ii} =  \sum_{a=\pm} \left\langle 2\mu_B^2 \mathcal{D}_{a,\bm{k}}
\tilde{g}^{(i)\,2}_{a,\bm{k}}\left\lbrace 1
    + \tilde{F}_{i,{\bm k}}^{(a)} \left[Y_a(\hat{k},T) -1\right] \right\rbrace \right\rangle_{a},\label{eq:Chi_Yosida_1}
\end{equation}
where  $\tilde{g}^{(i)}_{a,\bm{k}}$ is the effective $g$-factor for an
$i$-axis field, and 
$Y_a(\hat{k},T)$ is the angle-dependent Yosida function for the
intraband gap Eq. \ref{eq:FitA_1}. Explicit expressions for the
susceptibility in both the normal and superconducting states are
provided in the appendix~\ref{App:general}.
For even-parity gaps $\tilde{F}_{h,{\bm k}}^{(a)}=1$, and the pairing
suppresses the Pauli susceptibility, with it vanishing at zero
temperature.  On the other hand, an odd-parity state typically gives
only a partial suppression of the Pauli susceptibility; 
 In the extreme case
$\tilde{F}_{h,{\bm k}}^{(a)}=0$ the susceptibility is unaffected by the
superconductivity.

{\it $j=\frac{3}{2}$ superconductors}. As a concrete application of our approach we consider a system of
electrons with an effective 
$j=\frac{3}{2}$. A minimal model is given by
the Luttinger-Kohn Hamiltonian
\begin{equation}
\mathcal{H}_{0,\bm{k}}=\left(\alpha\left|\bm{k}\right|^2-\mu\right)\mathbb{1} +\beta_1\sum_{i}k_i^2J_i +\beta_2\sum_{i'\neq i}k_ik_{i'}J_iJ_{i'},\label{eq:LK_Ham}
\end{equation}
where $J_i$ are the spin-$\frac{3}{2}$ matrices and the indices $i$,
$i'$ run over Cartesian coordinates. The Hamiltonian can be cast into
the form of Eq. \ref{eq:H0} by defining the $\gamma$ matrices
$\vec{\gamma} = (\frac{1}{\sqrt{3}}[J_x^2-J_y^2], \frac{1}{3}[2J_z^2 -
J_x^2 - J_y^2], \frac{1}{\sqrt{3}} \{ J_y, J_z\} , \frac{1}{\sqrt{3}}
\{ J_x,J_y\} ,\frac{1}{\sqrt{3}} \{ J_x,J_z\} )$,  with
  corresponding coefficients $\vec{\varepsilon}=( \sqrt{3}\beta_1
(k_x^2-k_y^2)/2, \sqrt{3}\beta_1 (3k_z^2 -|\bm{k}|^2) / 2 ,
\sqrt{3}\beta_2 k_yk_z, \sqrt{3}\beta_2 k_xk_y, \sqrt{3}\beta_2 k_xk_z
)$. 
 The strong SOC in the Luttinger-Kohn model leads to a pronounced
spin-momentum locking.
This is most clearly observed in the limit
$\beta_1=\beta_2$, where the model has full spherical symmetry and the 
quantity ${\bm k}\cdot{\bf J}$ commutes with the Hamiltonian, i.e. the
projection of the spin along 
${\bm k}$ is a good quantum number. The eigenstates are classified
by the ``helical'' index
$\sigma$ such that $\hat{\bm k}\cdot{\bf J}|\sigma\rangle_{\bm k} =
\sigma|\sigma\rangle_{\bm k}$,
defining a spin-$\frac{3}{2}$ band ($\sigma = \pm\frac{3}{2}$) and 
spin-$\frac{1}{2}$ band ($\sigma = \pm\frac{1}{2}$).
This argument remains valid along
high-symmetry directions in the  presence of cubic anisotropy,
allowing us to  identify spin-$\frac{1}{2}$ and
-$\frac{3}{2}$ bands when $\beta_1\neq\beta_2$. As we will show
  below, compared to the spin-$\frac{1}{2}$-band, the
  spin-$\frac{3}{2}$ band exhibits counter-intuitive properties.

 The spin-momentum locking produces a highly anisotropic Zeeman
splitting of the bands by
an applied magnetic field $H_h=g\mu_B{\bm{h}}\cdot{\bf J}$. Although the general result using Eq.~\ref{eq:h_Fit} is complicated, in the 
rotationally-symmetric limit the effective $g$-factors take the
  compact forms
$\tilde{g}_{1/2}=\sqrt{1-3|\hat{\bm{h}}\cdot\hat{\bm{k}}|^2/4}$  and
$\tilde{g}_{3/2}=3|\hat{\bm{h}}\cdot\hat{\bm{k}}|/2$ in the
spin-$\frac{1}{2}$ and -$\frac{3}{2}$ bands, respectively; these
effective $g$-factors remain approximately valid in the general case.
Note that there is no splitting
of the spin-$\frac{3}{2}$ band in the direction perpendicular to the
field,  whereas the splitting of the spin-$\frac{1}{2}$ states is
maximal in this plane. 
To understand this, we
introduce the momentum-dependent angular 
momentum operators $J_z^{\bm k} = \hat{\bm k}\cdot{\bf J}$, and
raising and lowering operators $J_\pm^{\bm k}$ which satisfy $[J_z^{\bm k},{J}_{\pm}^{\bm k}]=\pm{J}_{\pm}^{\bm k}$.
The band
states are eigenstates of $J_z^{\bm k}$, i.e. $J_z^{\bm k} |\sigma\rangle_{\bm k} = \sigma|\sigma\rangle_{\bm
  k}$, and the action of the raising and lowering operators is $J_\pm^{\bm k}|\sigma\rangle_{\bm k} =
\sqrt{(\frac{3}{2}\mp\sigma)(\frac{3}{2}\pm\sigma +
  1)}|\sigma\pm1\rangle_{\bm k}$. Expressing the Zeeman Hamiltonian in terms of these operators we have
$H_h = g\mu_B \left({\bf h}\cdot\hat{\bm k}J_z^{\bm k} + h_+^{\bm
    k}J_+^{\bm k} + h_-^{\bm k}J_-^{\bm k}\right)$; the functions
$h_\pm^{\bm k} = -h_{\pm}^{-{\bm k}}$ are vanishing for $\hat{\bm
  k}\parallel{\bf h}$ and take maximal magnitude in the plane
perpendicular to ${\bm k}$; explicit expressions are given in the
appendix~\ref{App:J32}. Since the raising and lowering
operators do not couple the $\sigma=\pm\frac{3}{2}$ states, this immediately explains
why there is no splitting for ${\bm k}\perp {\bf
  h}$ in the spin-$\frac{3}{2}$ band; conversely, the larger matrix
elements for the raising and lowering operators compared to the
$z$-spin operator gives a maximum of the Zeeman splitting for the
spin-$\frac{1}{2}$ band in this plane.

 The spin-momentum locking also profoundly influences the
  superconducting gap structure, which we
illustrate by two odd-parity states: the fully-gapped 
$p$-wave triplet $A_{1u}$ state with $\tilde{\Delta}_{1u}=
\frac{2}{\sqrt{5}}({\bm{k}}\cdot{\bf J}) U_T$, and a
nodal $p$-wave septet $A_{2u}$ state $\tilde{\Delta}_{2u}=
\frac{1}{\sqrt{3}}({\bm{k}}\cdot{\bf T}) U_T$, where
$T_i = \{J_i, J_{i+1}^2-J_{i+2}^2\}$ for $i$ defined cyclically
(e.g. $J_{y+2}=J_x$). These two states are the leading $p$-wave instabilities
in the spin-$\frac{3}{2}$ band in the rotationally-symmetric
limit~\cite{Link2022}.  Using Eq. \ref{eq:FitA_1} we find that the
gaps opened by these pairing states have nontrivial band
dependence. This is most apparent in the $A_{2u}$ case, which in
the spin-$\frac{1}{2}$ band has fourteen nodes along the $(100)$ and
$(111)$ and equivalent directions, whereas in the spin-$\frac{3}{2}$
band it only has six 
nodes along the the $(100)$ directions~\cite{Venderbos_PRX_2018}.  
These nodes result
from the spin-momentum locking: along the $(100)$
directions the $A_{2u}$ state pairs electrons with helicity differing
by $\pm2$ which cannot be satisfied in either band, and hence implies
purely interband pairing and a node in the 
intraband gap. On the other hand, along the $(111)$ direction the
$A_{2u}$ state pairs electrons with helicity differing by $\pm3$, and
so it only opens a gap in the
spin-$\frac{3}{2}$ band. 
In contrast, the $A_{1u}$ state pairs electrons with the same
helicity, and hence it opens a full gap in both bands. Further
details on these pairing states are provided in the appendix~\ref{App:J32}. 

\begin{figure}[tb]
\begin{minipage}{0.49\columnwidth}
\includegraphics[width=\columnwidth]{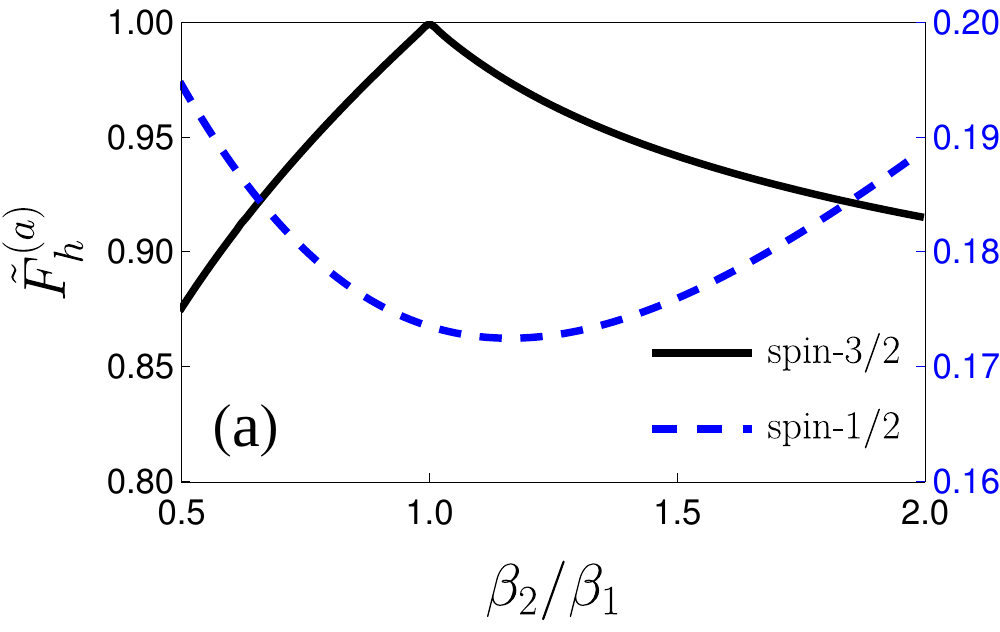}
\end{minipage}\hfil
\begin{minipage}{0.49\columnwidth}
\includegraphics[width=\columnwidth]{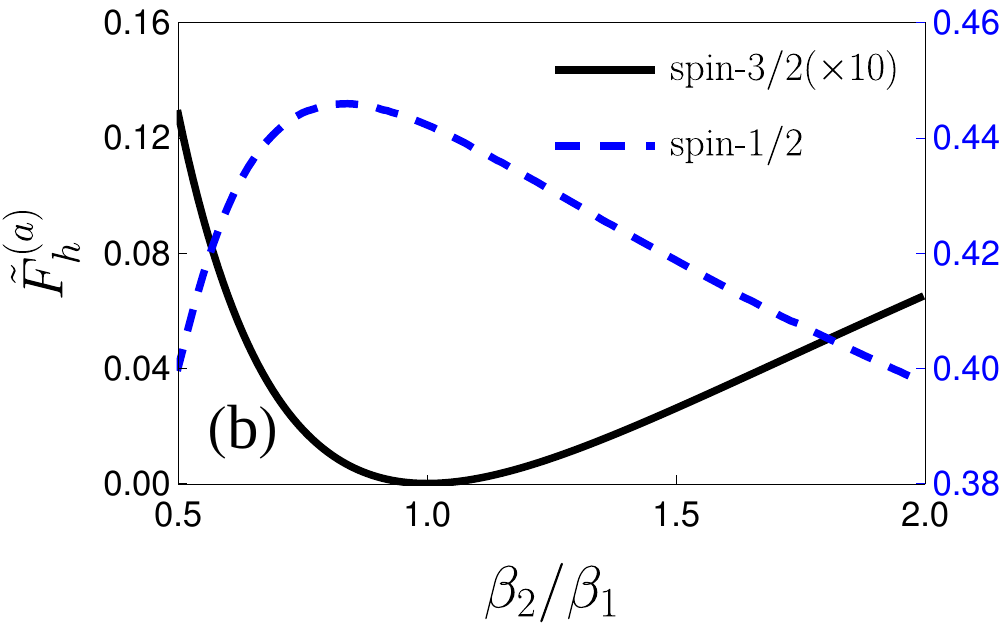}
\end{minipage}

\begin{minipage}{0.485\columnwidth}
\includegraphics[width=0.95\columnwidth]{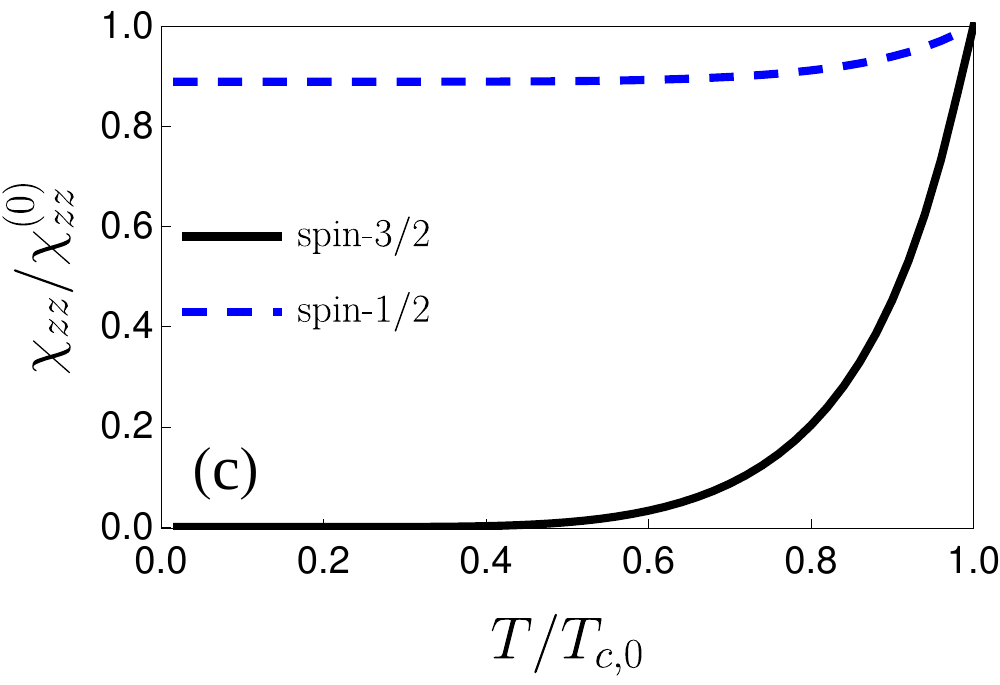}
\end{minipage}\hfil
\begin{minipage}{0.485\columnwidth}
\includegraphics[width=0.95\columnwidth]{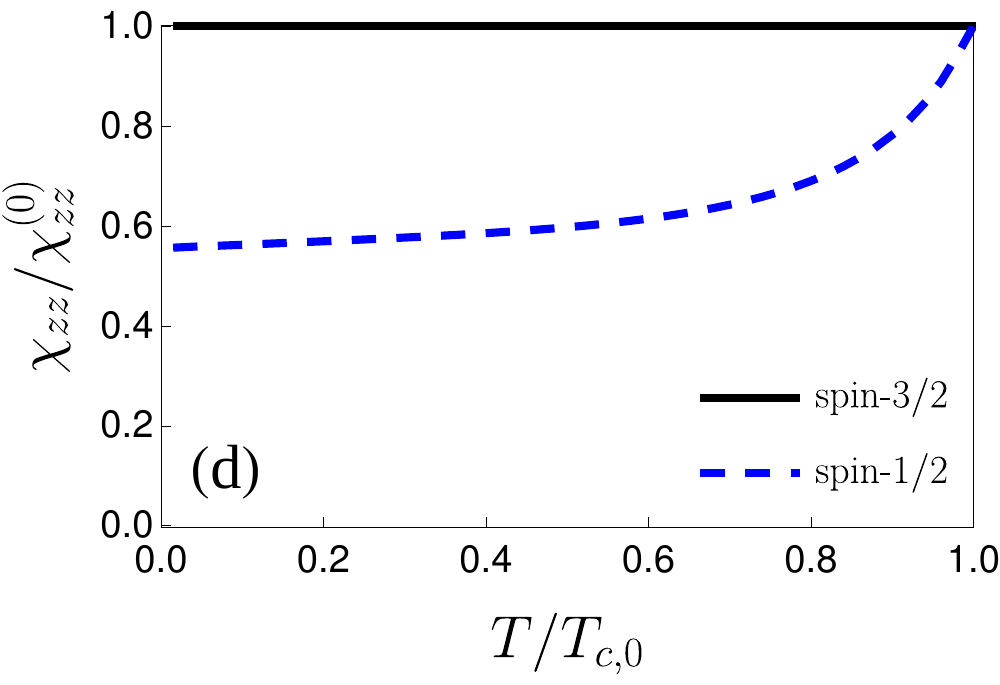}
\end{minipage}

\begin{minipage}{0.485\columnwidth}
\includegraphics[width=0.92\columnwidth]{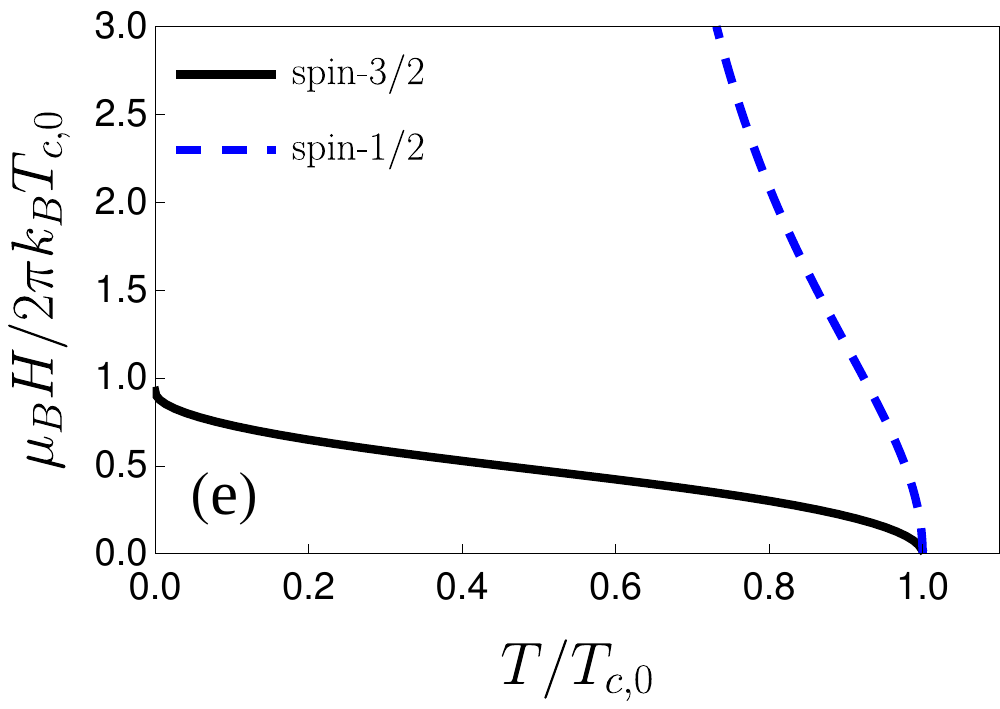}
\end{minipage}\hfil
\begin{minipage}{0.485\columnwidth}
\includegraphics[width=0.92\columnwidth]{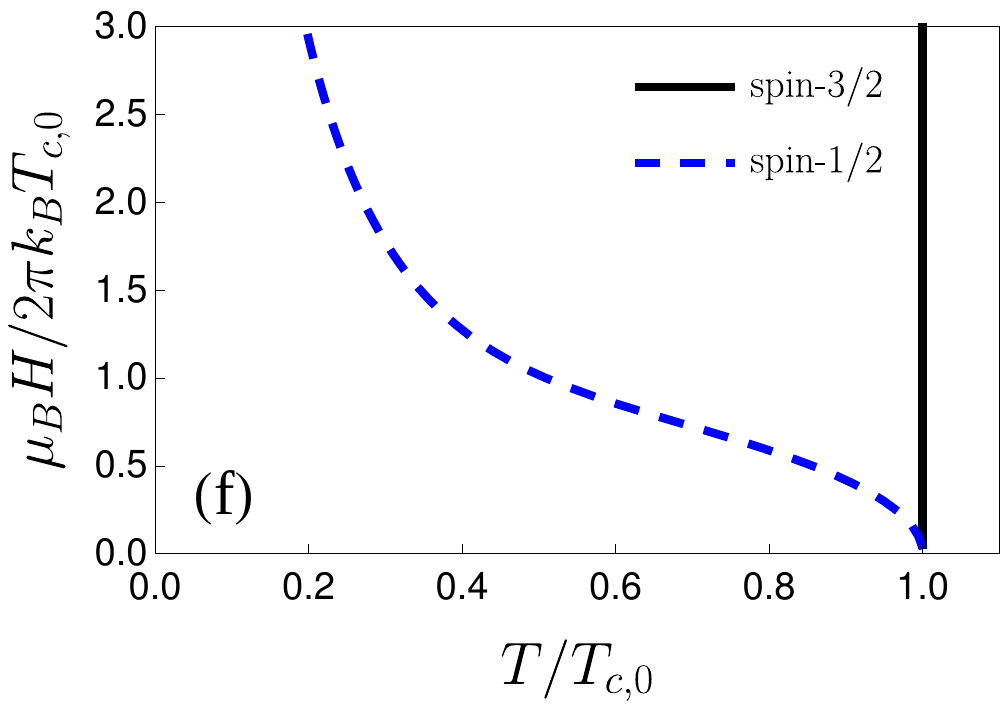}
\end{minipage}

	\caption{Comparison of $A_{1u}$ and $A_{2u}$ states. (a),
          (b): Normalized field-fitness
          $\tilde{F}_{h}^{(a)}$ on the spin-$3/2$ (black solid line,
          $\mu=-20$meV) and  spin-$1/2$  (blue dashed line,
          $\mu=20$meV) band Fermi surface for $A_{1u}$ (a) and
          $A_{2u}$ (b) pairing. (c), (d): Pauli susceptibility as a
          function of temperature at $\beta_2=\beta_1$ for $A_{1u}$
          (c) and $A_{2u}$ (d) pairing. (e), (f) Upper critical
          field excluding orbital effects for each band at
          $\beta_2=\beta_1$ for $A_{1u}$ (e) and $A_{2u}$ (f)
         pairing. In all plots we take $\alpha =20(a/\pi)^2$eV,
          $\beta_1=-15(a/\pi)^2$eV. } 
	\label{fig:spin_3on2_combined}
\end{figure}

The response of the superconductivity to an applied Zeeman field
displays major differences between the two bands. For a field applied
along the $z$-axis, the  averaged field-fitness function $\tilde{F}_{h}^{(a)} =
\langle \tilde{F}_{h,{\bm k}}^{(a)}\rangle_a$ is
plotted in Fig. \ref{fig:spin_3on2_combined}(a) and (b) as a function
of $\beta_2/\beta_1$ for the $A_{1u}$ and $A_{2u}$ states,
respectively. In the spin-$\frac{1}{2}$ band, the nonzero value of
$\tilde{F}_{h}^{(a)}$ indicates that a magnetic field is partially
pair-breaking for both states; in the spin-$\frac{3}{2}$ band,
however, the field is almost completely pair-breaking for the $A_{1u}$
state, whereas the $A_{2u}$ state is almost insensitive to its effects. These
results are only weakly dependent upon the ratio $\beta_2/\beta_1$,
and become exact in the rotationally-symmetric limit. This 
behaviour is reflected in the susceptibility, as shown in 
Fig. \ref{fig:spin_3on2_combined}(c) and (d): in the
spin-$\frac{3}{2}$ band the Pauli contribution
to the susceptibility is completely suppressed for the $A_{1u}$ state,
whereas we see no change in the susceptibility for the $A_{2u}$
state. In contrast, in both cases there is moderate suppression of the
susceptibility in the spin-$\frac{1}{2}$ band. Likewise, as shown in
Fig.~\ref{fig:spin_3on2_combined}(e) and (f), in the
spin-$\frac{3}{2}$ band the $A_{1u}$ state is Pauli-limited whereas
the $A_{2u}$ state is only limited by orbital effects.

The remarkable response to a magnetic field can be understood in the
rotationally-symmetric limit in terms
of the helicity quantum number. Projected onto the 
spin-$\frac{3}{2}$ band, the $A_{1u}$ state pairs 
electrons with the same helicity whereas the $A_{2u}$ state pairs
electrons with opposite helicity. In the projection of the Zeeman
Hamiltonian only the $J^{z}_{\bm k}$ operator has nonzero matrix
elements, and so the energy shift of the state $|\sigma\rangle_{\bm
  k}$ is $\sigma {\bf h}\cdot\hat{\bm k}$. Since the $A_{2u}$ state 
pairs electrons which have the same Zeeman shift, it does not
experience pair-breaking; in contrast, the $A_{1u}$
state pairs electrons with opposite Zeeman shift, and the
pair-breaking is maximal. The situation in the
spin-$\frac{1}{2}$ band is more complicated, since helicity is not a
good quantum number for the projected Zeeman Hamiltonian, and the
$A_{2u}$ state has both equal- and opposite-helicity pairing matrix
elements. This gives the intermediate values of $\tilde{F}_h^{(a)}$ for
the spin-$\frac{1}{2}$ band. Although this argument is only rigorously
valid in the rotationally-symmetric limit, the plot of
$\tilde{F}_{h}^{(a)}$ in Fig. \ref{fig:spin_3on2_combined}(a) and (b)
indicates that it remains valid more generally.

{\it Conclusions}. In this paper we have developed a basis-independent framework to understand
the interplay of superconductivity and symmetry-breaking perturbations
in a multiband system. Using a generic minimal model, we have shown
that that coupling of orbital and spin degrees of freedom typically
reduces the effect of the perturbation via the appearance of an
effective $g$-factor Eq.~\ref{eq:h_Fit}. Moreover, the pair-breaking
effect of this perturbation can be formulated in terms of the
field-fitness function Eq.~\ref{eq:FieldFitnessFunction_1}. Together, these
quantities control the suppression of the critical temperature by 
the perturbation, and in the case of a Zeeman field determine the
magnetic susceptibility below the critical temperature. To
illustrate these effects, we have examined $p$-wave pairing of
effective spin-$\frac{3}{2}$ electrons with normal state described by
the Luttinger-Kohn model. The characteristic spin-momentum locking in
this model leads to the remarkable result that in the
spin-$\frac{3}{2}$ band the triplet $A_{1u}$ state is strongly
suppressed by a Zeeman field, whereas the septet $A_{2u}$ state is
largely immune to it, giving dramatically different behaviour in
the upper critical field and the spin susceptibility.

{\it Acknowledgements} D.C.C. and P.M.R.B were supported by
the Marsden Fund Council from Government funding, managed by Royal
Society Te Ap\={a}rangi. D.F.A. was supported 
by the U.S. Department of Energy, Office of Basic Energy
Sciences, Division of Materials Sciences and Engineering,
under Award No. DE-SC0021971.

\bibliography{Hc2_references}

\appendix
\section{General results on the effect of broken symmetry} \label{App:general}

\subsection{The Green's function}\label{App:projection}

In the absence of symmetry-breaking perturbations, the Green's function is
\begin{equation}
G\left(\bm{k},i\omega_n\right)=\sum_{a = \pm} \frac{1}{i\omega_n-\xi_{a,\bm{k}}} \mathcal{P}_{a,\bm{k}},
\end{equation}
 where $\mathcal{P}_{a,\bm{k}}$ projects onto band $a$ at
wavevector ${\bm k}$, given explicitly by
\begin{equation}
\mathcal{P}_{a,\bm{k}} = \frac{\mathbb{1} + a \hat{\varepsilon}_{\bm{k}}\cdot\vec{\gamma}}{2}= \frac{1}{2}\left[\mathbb{1}+a\frac{\bar{\mathcal{H}}_{0,\bm{k}}}{\left|\vec{\varepsilon}_{\bm{k}}\right|}\right].
\end{equation}

 We now consider the addition of the symmetry-breaking
perturbation. Assuming that $h_{\bm k}\ll |\vec{\epsilon}_{\bm k}|$,
i.e. the mixing between the unperturbed bands is negligible, we can
approximate the Green's function by 
\begin{equation}
G\left(\bm{k},i\omega_n\right) =\sum_{a,b = \pm} \frac{1}{i\omega_n-\xi_{a,b,\bm{k}}} \mathcal{P}_{a,b,\bm{k}},\label{eq:Gab}
\end{equation}
where $\xi_{a,b,{\bm k}} = \xi_{a,{\bm k}} + b\tilde{g}_{a,{\bm k}}h_{\bm
  k}$
and the projection operator is defined
\begin{equation}
\mathcal{P}_{a,b,\bm{k}} =
\frac{1}{4}\left(\mathbb{1}+a\frac{\bar{\mathcal{H}}_{0,\bm{k}}}{\left|\vec{\varepsilon}_{\bm{k}}\right|}\right)\left(\mathbb{1}
  +ab\frac{\left\lbrace \bar{\mathcal{H}}_{0,\bm{k}} , \mathcal{H}_{h,
        \bm{k}}
    \right\rbrace}{2\tilde{g}_{a,\bm{k}}h\left|\vec{\varepsilon}_{\bm{k}}\right|}\right) 
\end{equation}
This expression is valid to lowest order in
$h_{\alpha,\bm{k}}/|\vec{\varepsilon}_{\bm{k}}|$, which is equivalent
to ignoring interband corrections. 

\subsection{The linearized gap equation}\label{App:linearized}

The pairing potential $\Delta_0$ is determined by self-consistent
solution of the gap
equation
\begin{equation}
\Delta_0=-V_0\beta \sum_{i\omega_n,\bm{k}}
\text{Tr}\{\tilde{\Delta}^\dagger_{\bm k}F\left(\bm{k},i\omega_n\right)\},\label{eq:appGapEq}
\end{equation}
where $F({\bm k},i\omega_n)$ is the anomalous Green's function and
$V_0$ is the pairing interaction. Ignoring interband pairing, just
below the critical temperature the anomalous Green's function in the
presence of symmetry-breaking perturbations can be approximated by 
\begin{equation}
F\left(\bm{k},i\omega_n\right) = \sum_{a,b,b'}
\frac{\mathcal{P}_{a,b,\bm{k}}\Delta_{\bm k} \mathcal{P}_{a,b',-\bm{k}}^T}{\left(i\omega_n - \xi_{a,b,\bm{k}}\right)\left(i\omega_n +
    \xi_{a,b',\bm{k}}
  \right)}\,.
\end{equation}
Inserting this into the gap equation we obtained the linearized form
\begin{equation}
-\frac{1}{V_0} = \beta \sum_{i\omega_n,\bm{k}}\sum_{a,b,b'}
\frac{\text{Tr}\lbrace \tilde{\Delta}_{\bm
    k}^\dagger\mathcal{P}_{a,b,\bm{k}}\tilde{\Delta}_{\bm k} \mathcal{P}_{a,\bar{b}',\bm{k}}\rbrace}{\left(i\omega_n - \xi_{a,b,\bm{k}}\right)\left(i\omega_n + \xi_{a,b',\bm{k}}\right)}\,.
\end{equation}
Performing the sums over Matsubara frequency and momentum gives Eq. \ref{eq:logTc_h_1} after the zero-field result is subtracted.

\subsection{Spin susceptibility}\label{App:susceptibility}

The static long-wavelength paramagnetic $i$-axis susceptibility can be divided
into an intraband Pauli contribution and an interband van Vleck
contribution 
\begin{equation}
\chi_{ii} = \sum_{a}\{\chi_{ii,aa} + \chi_{ii,a\bar{a}}\}\,.
\end{equation}
In the normal state these terms are given by 
\begin{align}
	\chi_{ii,aa}^{(N)} &=\sum_{\bm{k}} \frac{\mu_B^2\bar{\chi}_{ii,aa}^{(N)}\left(\bm{k}\right)}{8\left|\vec{\varepsilon}_{\bm{k}}\right|^2}\text{Tr}\lbrace |\left\lbrace {S}_{i} , \bar{\mathcal{H}}_{0,\bm{k}}\rbrace|^2\mathcal{P}_{a,\bm{k}}\right\rbrace \\
	\chi_{ii,a\bar{a}}^{(N)} &=\sum_{\bm{k}}  \frac{\mu_B^2\bar{\chi}_{ii,a\bar{a}}^{(N)}\left(\bm{k}\right)}{16\left|\vec{\varepsilon}_{\bm{k}}\right|^2}\text{Tr}\lbrace|[ {S}_{i} , \bar{\mathcal{H}}_{0,\bm{k}}]|^2 \rbrace\label{eq:NormChiDefn}.
	\end{align}
Here ${S}_i$ is the spin operator in the $i$ direction. In 
spin-$\frac{1}{2}$ materials this is given by
${S}_i=\tau_0\otimes\sigma_i$, whereas  ${S}_i=J_i$ for effective
spin-$\frac{3}{2}$ materials. The traces in these expression encode
the effect of the spin-orbital texture of the band electron states;
note the appearance of the effective $g$-factor in the Pauli
contribution. This modulates the Lindhard function
	\begin{align}
	\bar{\chi}_{ii,aa'}^{(N)}\left(\bm{k}\right)&= \lim\limits_{\bm{q}\rightarrow\bm{0}}\frac{n_{F}\left(\xi_{a,\bm{k}+\bm{q}}\right)-n_{F}\left(\xi_{a',\bm{k}}\right)}{\xi_{a,\bm{k}+\bm{q}}-\xi_{a',\bm{k}}}
	\end{align}
where $n_F$ is the Fermi-Dirac function.

The Pauli susceptibility is significantly modified below the critical
temperature due to the opening of the superconducting gap. Moreover,
it develops an anomalous contribution due to the pairing
\begin{widetext}
	\begin{align}
	\chi_{ii,aa}^{(A)} &= \sum_{\bm{k}}
                             \frac{\mu_B^2\bar{\chi}_{ii,aa}^{(A)}\left(\bm{k}\right)}{256\left|\vec{\varepsilon}_{\bm{k}}\right|^4}\text{Tr}\lbrace
                             (|\lbrace \lbrace {S}_{i} ,
                             \bar{\mathcal{H}}_{0,\bm{k}}\rbrace ,
                             \lbrace \tilde{\Delta}_{\bm k} ,
                             \bar{\mathcal{H}}_{0,\bm{k}}\rbrace
                             \rbrace|^2 -|[ \lbrace {S}_{i} ,
                             \bar{\mathcal{H}}_{0,\bm{k}}\rbrace ,
                             \lbrace \tilde{\Delta}_{\bm k} , \bar{\mathcal{H}}_{0,\bm{k}}\rbrace]|^2 )\mathcal{P}_{a,\bm{k}} \rbrace.\label{eq:AnomChiDefn}%
	\end{align}
The modified Lindhard function appearing in this expression is  
	\begin{align}
	\bar{\chi}_{ii,aa'}^{(A)}\left(\bm{k}\right)&= \lim\limits_{\bm{q}\rightarrow\bm{0}}\frac{\left|\Delta_{0,\bm{k}+\bm{q}}\right|\left|\Delta_{0,\bm{k}}\right|}{E_{a,\bm{k}+\bm{q}}E_{a',\bm{k}}}\left\lbrace \frac{n_{F}\left(E_{a,\bm{k}+\bm{q}}\right)-n_{F}\left(E_{a',\bm{k}}\right)}{E_{a,\bm{k}+\bm{q}}-E_{a',\bm{k}}}-\frac{1-n_{F}\left(E_{a,\bm{k}+\bm{q}}\right)-n_{F}\left(E_{a',\bm{k}}\right)}{E_{a,\bm{k}+\bm{q}}+E_{a',\bm{k}}}  \right\rbrace,
	\end{align}
where $E_{a,{\bm k}} = \sqrt{\xi_{a,{\bm k}}^2 + |\Delta_{a,{\bm k}}|^2}$.
The Lindhard function appearing in the ``normal'' contribution is also
modified as
	\begin{align}
	\bar{\chi}_{ii,aa'}^{(N)}\left(\bm{k}\right)&= \lim\limits_{\bm{q}\rightarrow\bm{0}}\frac{-\mu_B }{2}\left\lbrace \left[1+\frac{\xi_{a,\bm{k}+\bm{q}}\xi_{a',\bm{k}}}{E_{a,\bm{k}+\bm{q}}E_{a',\bm{k}}}\right]\frac{n_{F}\left(E_{a,\bm{k}+\bm{q}}\right)-n_{F}\left(E_{a',\bm{k}}\right)}{E_{a,\bm{k}+\bm{q}}-E_{a',\bm{k}}}\right.\nonumber\\
	&\qquad\qquad\qquad\left. + \left[1-\frac{\xi_{a,\bm{k}+\bm{q}}\xi_{a',\bm{k}}}{E_{a,\bm{k}+\bm{q}}E_{a',\bm{k}}}\right]\frac{1-n_{F}\left(E_{a,\bm{k}+\bm{q}}\right)-n_{F}\left(E_{a',\bm{k}}\right)}{E_{a,\bm{k}+\bm{q}}+E_{a',\bm{k}}} \right\rbrace,
	\end{align}
\end{widetext}
If the superconducting gap has even parity, the anomalous contribution to the Pauli susceptibility will exactly cancel the normal part as $T\rightarrow 0$. More generally any such cancellation depends sensitively on the spin-orbital texture and structure of the gap. 

\subsection{Broken inversion symmetry}\label{App:ISB}

In the main text we focus on the effect of a perturbation which breaks
time-reversal symmetry. Our formalism, can also account for inversion
symmetry-breaking perturbations such as a weak antisymmetric spin-orbit coupling. 
In the presence of an inversion symmetry breaking perturbation, the
critical temperature is given by Eq. \ref{eq:logTc_h_1}, with the
field-fitness replaced by 
\begin{equation}
\tilde{F}_{h,{\bm k}}^{(a)}=\frac{\text{Tr}\lbrace
    |[\lbrace\bar{\mathcal{H}}_{0,\bm{k}},\tilde{\Delta}_{\bm
          k}\rbrace,\lbrace
        \bar{\mathcal{H}}_{0,\bm{k}},\mathcal{H}_{h,\bm{k}}\rbrace]|^2
        \mathcal{P}_{a,{\bm k}}\rbrace}{2\text{Tr}\lbrace |\lbrace
        \bar{\mathcal{H}}_{0,\bm{k}},\mathcal{H}_{h,\bm{k}}\rbrace|^2\mathcal{P}_{a,{\bm k}}
        \rbrace\text{Tr}\lbrace |\lbrace
        \bar{\mathcal{H}}_{0,\bm{k}},\tilde{\Delta}_{\bm
          k}\rbrace|^2\mathcal{P}_{a,{\bm k}} \rbrace}. \label{eq:FieldFitnessFunctions_Gen}
\end{equation}
The key difference compared to Eq.~\ref{eq:FieldFitnessFunction_1} of
the main text is that the numerator involves the commutator of
the anticommutators, as opposed to the anticommutator of the
anticommutators. It can be straightforwardly shown that
$\tilde{F}_{h,{\bm k}}^{(a)}=0$ for all even-parity superconductors,
reflecting the fact that the intraband potential of these states
always pairs time-reversed partners.

\section{$J=\frac{3}{2}$ superconductors}\label{App:J32}

\subsection{Spherical limit}

In the spherical limit the Luttinger-Kohn Hamiltonian takes the form
\begin{equation}
H  =  \alpha |{\bm{k}}|^2\hat{1}_4 + \beta ({\bm{k}}\cdot {\bf J})^2\,.
\end{equation}
We observe that the Hamiltonian has two doubly-degenerate bands
corresponding to $\langle {\bm{k}}\cdot {\bf J} \rangle =
\pm\frac{3}{2}|{\bm{k}}|$  and 
$\langle {\bm{k}}\cdot {\bf J} \rangle = \pm\frac{1}{2}|{\bm
  k}|$. We will
refer to these two bands as the spin-$\frac{3}{2}$ and spin-$\frac{1}{2}$
bands, respectively. A natural choice of basis for these bands is the
``helical'' basis
\begin{equation}
\{|\tfrac{3}{2}\rangle_{\bm{k}}\,,\,\, |-\tfrac{3}{2}\rangle_{\bm{k}}\}\,,
\qquad \{|\tfrac{1}{2}\rangle_{\bm{k}}\,,\,\, |-\tfrac{1}{2}\rangle_{\bm{k}}\}
\end{equation}
where ${\bm{k}}\cdot{\bf J}|\sigma\rangle_{\bm{k}} = \sigma|{\bm{k}}||\sigma\rangle_{\bm{k}}$. Note that the operator ${\bm{k}}\cdot\hat{\bf
  J}$ is invariant under time-reversal, and so the ``helicity'' $\sigma$ is \emph{not} a
pseudospin index. Expressed in terms of the $z$-spin eigenstates we
have
\begin{eqnarray}
|\tfrac{3}{2}\rangle_{\bm{k}} &=&
e^{-3i\phi}\cos^3(\tfrac{\theta}{2})|\tfrac{3}{2}\rangle
+\sqrt{3}e^{-2i\phi}\cos^2(\tfrac{\theta}{2})\sin(\tfrac{\theta}{2})|\tfrac{1}{2}\rangle
                                 \notag \\
&& +\sqrt{3}e^{-i\phi}\cos(\tfrac{\theta}{2})\sin^2(\tfrac{\theta}{2})|-\tfrac{1}{2}\rangle  
+\sin^3(\tfrac{\theta}{2})|-\tfrac{3}{2}\rangle \notag \\\\
|\tfrac{1}{2}\rangle_{\bm{k}} &=&
-\sqrt{3}e^{-3i\phi}\sin(\tfrac{\theta}{2})\cos^2(\tfrac{\theta}{2})|\tfrac{3}{2}\rangle
                                 \notag \\ &&+\frac{1}{2}e^{-2i\phi}[-1+3\cos(\theta)]\cos(\tfrac{\theta}{2})|\tfrac{1}{2}\rangle 
\notag \\ && +\frac{1}{2}e^{-i\phi}[1+3\cos(\theta)]\sin(\tfrac{\theta}{2})|-\tfrac{1}{2}\rangle
\notag \\
&&
+\sqrt{3}\sin^2(\tfrac{\theta}{2})\cos(\tfrac{\theta}{2})|-\tfrac{3}{2}\rangle
\\
|-\tfrac{1}{2}\rangle_{\bm{k}} &=&
\sqrt{3}e^{-3i\phi}\sin^2(\tfrac{\theta}{2})\cos(\tfrac{\theta}{2})|\tfrac{3}{2}\rangle
\notag \\ && -\frac{1}{2}e^{-2i\phi}[1+3\cos(\theta)]\sin(\tfrac{\theta}{2})|\tfrac{1}{2}\rangle 
\notag \\ &&
+\frac{1}{2}e^{-i\phi}[-1+3\cos(\theta)]\cos(\tfrac{\theta}{2})|-\tfrac{1}{2}\rangle
\notag \\
&&
+\sqrt{3}\sin(\tfrac{\theta}{2})\cos^2(\tfrac{\theta}{2})|-\tfrac{3}{2}\rangle\\
|-\tfrac{3}{2}\rangle_{\bm{k}} &=&
-e^{-3i\phi}\sin^3(\tfrac{\theta}{2})|\tfrac{3}{2}\rangle
+\sqrt{3}e^{-2i\phi}\sin^2(\tfrac{\theta}{2})\cos(\tfrac{\theta}{2})|\tfrac{1}{2}\rangle 
\notag \\ && -\sqrt{3}e^{-i\phi}\sin(\tfrac{\theta}{2})\cos^2(\tfrac{\theta}{2})|-\tfrac{1}{2}\rangle  
+\cos^3(\tfrac{\theta}{2})|-\tfrac{3}{2}\rangle \notag \\
\end{eqnarray}
where $(\phi,\theta)$ are the angular coordinates defined by the
direction of ${\bm{k}}$.
The helicity index reverses under inversion ${\cal I}$ but is
preserved under time-reversal ${\cal T}$:
\begin{gather}
{\cal I}|\pm \tfrac{3}{2}\rangle_{\bm{k}} = |\mp \tfrac{3}{2}\rangle_{\bf
  -k}\,, \quad {\cal I}|\pm \tfrac{1}{2}\rangle_{\bm{k}} = |\mp \tfrac{1}{2}\rangle_{\bf
  -k}\,, \\
{\cal T}|\pm\tfrac{3}{2}\rangle_{\bm{k}} = \mp e^{3i\phi}|\pm \tfrac{3}{2}\rangle_{\bf
  -k}\,, \quad {\cal T}|\pm\tfrac{1}{2}\rangle_{\bm{k}} = \pm e^{3i\phi}|\pm\tfrac{1}{2}\rangle_{\bf
  -k}\,,
\end{gather}

Because of the momentum-dependent quantization axis, it is useful to
define the momentum-dependent angular-momentum operators
\begin{equation}
J^{{\bm{k}}}_z =  \hat{\bf
  k}\cdot{\bf J}\,,\quad
J_{\pm}^{\bm{k}} = \left(\hat{\bm{\theta}}_{\bm{k}} \pm i\hat{\bm{\phi}}_{\bm{k}}\right)\cdot {\bf J} 
\end{equation}
where $\hat{\bm{\theta}}_{\bm{k}} =
\cos\phi\cos\theta\hat{\bf x} + \sin\phi\cos\theta\hat{\bf y} -
\sin\theta\hat{\bf z}$ and $\hat{\bm{\phi}}_{\bm{k}} =
-\sin\phi\hat{\bf x} + \cos\phi\hat{\bf y}$ are the canonical
spherical-coordinate unit vectors in
the directions perpendicular to $\hat{\bm{k}}$.

\subsection{The Zeeman field}

We include a Zeeman field in Hamiltonian as
\begin{equation}
H_{\text{Z}} =  g \mu_B {\bf h}\cdot{\bf J} \notag 
\end{equation}
To understand the effect of the spin-orbit coupling we now re-express
this in terms of the momentum-dependent angular momentum operators:
\begin{eqnarray}
H_{\text{Z}}^{\bm{k}} & = & g\mu_B h\left[\hat{\bf h}\cdot\hat{\bm
                           k}J_z^{\bm{k}} + \hat{\bf
                           h}\cdot\left(\hat{\bm{\theta}}_{\bm{k}} +
                           i\hat{\bm{\phi}}_{\bm{k}}\right)J_+^{\bm k}
                           \right. \notag \\
&& \left. + \hat{\bf
                           h}\cdot\left(\hat{\bm{\theta}}_{\bm{k}} -
                           i\hat{\bm{\phi}}_{\bm{k}}\right)J_-^{\bm k}
\right] \,.\label{eq:Hzk}
\end{eqnarray}
Note that for ${\bm k}$ in the
direction of the applied field, only the first term is present, and
the Zeeman field is diagonal in the helical basis. Conversely, for ${\bm k}$ perpendicular to the applied field, only the
second and third terms are present, and the Zeeman field couples the
helical states with helicity
$\sigma$ differing by $\pm 1$.

\subsection{The pairing states}

\subsubsection{The $p$-wave $A_{1u}$ state}

The $p$-wave $A_{1u}$ has
\begin{equation}
\tilde{\Delta}({\bm{k}}) = {\bm{k}}\cdot{\bf J} =
|{\bm{k}}|J_z^{\bm{k}}
\end{equation}
Since $\tilde{\Delta}({\bm{k}})$ is proportional to $J_z^{\bm{k}}$, and
the helicity is preserved under time-reversal, this pairs states with
the same 
helicity. This implies that in the spherical
limit the $A_{1u}$ state corresponds to purely intraband pairing.

\subsubsection{The $p$-wave $A_{2u}$ state}

The $p$-wave $A_{2u}$ has the pair potential
\begin{widetext}
\begin{eqnarray}
\tilde{\Delta}({\bm{k}}) &=& 
\sum_{i=x,y,z} k_i[J_{i+1}J_iJ_{i+1} -J_{i+2}J_iJ_{i+2}]\notag \\
& = & |{\bm{k}}|\left(\frac{1}{32}[6\cos2\phi\sin2\theta
  + 3i(3+\cos2\theta)\sin\theta\sin2\phi]J^{\bm
    k}_{+}J^{\bm{k}}_{+}J^{\bm{k}}_+ \right. \notag \\
&& + \frac{1}{32}[6\cos2\phi\sin2\theta
  - 3i(3+\cos2\theta)\sin\theta\sin2\phi]J^{\bm
    k}_{-}J^{\bm{k}}_{-}J^{\bm{k}}_- \notag \\
&& + \frac{1}{8}[4\cos2\phi\cos2\theta
  + i(1+3\cos2\theta)\cos\theta\sin2\phi]J^{\bm
    k}_{+}J^{\bm{k}}_{z}J^{\bm{k}}_+ \notag \\
&& + \frac{1}{8}[4\cos2\phi\cos2\theta
  - i(1+3\cos2\theta)\cos\theta\sin2\phi]J^{\bm
    k}_{-}J^{\bm{k}}_{z}J^{\bm{k}}_- \notag \\
&& -\frac{1}{128}\left[2\cos2\phi\sin2\theta
  + i(1+3\cos2\theta)\sin\theta\sin2\phi\right](7J^{\bm{k}}_+ - 20 J^{\bm
    k}_{z}J^{\bm{k}}_{+}J^{\bm{k}}_z) \notag \\
&& \left. -\frac{1}{128}\left[2\cos2\phi\sin2\theta
  - i(1+3\cos2\theta)\sin\theta\sin2\phi\right](7J^{\bm{k}}_- - 20 J^{\bm
    k}_{z}J^{\bm{k}}_{-}J^{\bm{k}}_z)\right)
\end{eqnarray}
\end{widetext}
Although very complicated compared to the $A_{1u}$ case, we observe
that the pair potential never pairs electrons with the same helicity,
since it always involves operators which change the helicity. That is,
projected into the band basis, the pair potential only pairs the
states $|\sigma\rangle_{\bm{k}}$ and $|-\sigma\rangle_{-{\bf
    k}}$. Moreover, we have the following special cases:
\begin{itemize}
\item along the crystal axes
$(\phi,\theta) = (0,\frac{\pi}{2}),$ $(\frac{\pi}{2},\frac{\pi}{2})$,
$(\phi,0)$, only the coefficients of $J_\pm^{\bm{k}}J_z^{\bm
  k}J_{\pm}^{\bm k}$ are nonzero, i.e. the pairing potential pairs
states with helicity differing by $\pm 2$. This cannot be satisfied
in either band, and so the intraband gap develops a node.
\item along the crystal body diagonals
$(\phi,\theta) = (\pm \frac{\pi}{4},\pm\arctan\sqrt{2})$ only the
coefficients of $J_\pm^{\bm{k}}J_\pm^{\bm
  k}J_{\pm}^{\bm k}$ are nonzero, i.e. the pairing potential pairs
states with helicity differing by $\pm 3$, i.e. there is only pairing
in the spin-$\frac{3}{2}$ band. Consequently the gap in the
spin-$\frac{1}{2}$ band has nodes.
\end{itemize}

\end{document}